\documentclass[aps,prb,twocolumn,superscriptaddress, amsfonts,amssymb,amsmath,titlepage]{revtex4-2}

\usepackage[hidelinks, breaklinks=true, colorlinks=true, allcolors=cyan]{hyperref}
\usepackage{latexsym}
\usepackage{graphicx}
\usepackage{textcomp}
\usepackage{xcolor}
\usepackage{multirow} 
\usepackage[space]{grffile} 
\usepackage{xspace} 
\usepackage{ulem}   
\usepackage[product-units=power, separate-uncertainty=true, multi-part-units=single]{siunitx}
\usepackage{currfile}
\usepackage{CJKutf8} 

\newcommand{\pgi}{Peter Gr\"{u}nberg Institut (PGI-3), Forschungszentrum J\"{u}lich, 52425 J\"{u}lich, Germany}
\newcommand{\jara}{J\"{u}lich Aachen Research Alliance (JARA), Fundamentals of Future Information Technology, 52425 J\"{u}lich, Germany}
\newcommand{\RWTH}{Experimentalphysik IV A, RWTH Aachen University, Otto-Blumenthal-Strasse, 52074 Aachen, Germany}
\newcommand{\diam}{Diamond Light Source Ltd, Didcot, OX110DE, Oxfordshire, United Kingdom}
\newcommand{\duke}{Thomas Lord Department of Mechanical Engineering and Materials Science, Duke University, Durham, NC 27708, USA}
\newcommand{\etal}{\textit{et al.}\xspace}
\newcommand{\myref}[7]{#1, #2, \href{http://dx.doi.org/#7}{#3 \textbf{#4}, #5 (#6).}} 
\newcommand{\tB}{$30^\circ$-tBLG\xspace}
\newcommand{\Gnul}{G-$R0^\circ$\xspace}
\newcommand{\lqGnul}{lqG-$R0^\circ$\xspace}

\newcommand{\BN}{B$_x$N$_y$\xspace}
\newcommand{\BNnul}{B$_x$N$_y$-$R0^\circ$\xspace}
\newcommand{\BBN}{B$_{\mathrm{B}_x\mathrm{N}_y}$\xspace}
\newcommand{\NBN}{N\xspace}
\newcommand{\BZL}{B$_\mathrm{ZL}$\xspace}
\newcommand{\BSiC}{B$_\mathrm{SiC}$\xspace}
\newcommand{\CSiC}{C$_\mathrm{SiC}$\xspace}
\newcommand{\SiSiC}{Si$_\mathrm{SiC}$\xspace}

\newcommand{\CGnul}{C$_{\mathrm{G}}$\xspace}

\newcommand{\C}{$^\circ$C\xspace}

\newcommand{\blue}[1]{#1}
\newcommand{\BZ}{$\%\mathrm{BZ_{SiC}}$\xspace}

\newcommand{\Pc}{$P_\mathrm{c}^\textbf{H}$\xspace}
\newcommand{\Fc}{$F_\mathrm{c}^\textbf{H}$\xspace}
\newcommand{\three}{$\left(3\times3\right)$\xspace}

\newcommand{\seven}{$\left(7\times7\right)$\xspace}
\newcommand{\sqthree}{$\left(\sqrt{3}\times\sqrt{3}\right)\mathrm{R}\ang{30}$\xspace}

\newcommand{\borazine}{$\mathrm{B}_{3}\mathrm{H}_{6}\mathrm{N}_{3}$\xspace}

\newsavebox{\smlmat}
\savebox{\smlmat}{$\left(\smallmatrix 3&~1 \\ -1&~2 \endsmallmatrix \right)$\xspace}
\newcommand{\extraspots}{\usebox{\smlmat}}

\begin{document}

\title{Boron nitride on SiC(0001)}
\author{You-Ron~Lin (\begin{CJK*}{UTF8}{bsmi}林又容\end{CJK*})}   
								\affiliation{\pgi} \affiliation{\jara} \affiliation{\RWTH}
\author{Markus~Franke}      	\affiliation{\pgi} \affiliation{\jara}
\author{Shayan~Parhizkar}	\email{Present address: Chair of Electronic Devices, RWTH Aachen University, Otto-Blumenthal-Str. 2, 52074 Aachen, Germany} \affiliation{\pgi} \affiliation{\jara}
\author{Miriam~Raths}       	\affiliation{\pgi} \affiliation{\jara} \affiliation{\RWTH}
\author{Victor~Wen-zhe~Yu}	\email{Present address: Materials Science Division, Argonne National Laboratory, Lemont, Illinois 60439, USA}       \affiliation{\duke}
\author{Tien-Lin~Lee (\begin{CJK*}{UTF8}{bsmi}李天麟\end{CJK*})}   
								\affiliation{\diam}
\author{Serguei~Soubatch}    	\affiliation{\pgi} \affiliation{\jara}
\author{Volker~Blum}    	\affiliation{\duke}
\author{F.~Stefan~Tautz}    	\affiliation{\pgi} \affiliation{\jara} \affiliation{\RWTH}
\author{Christian~Kumpf}	\affiliation{\pgi} \affiliation{\jara} \affiliation{\RWTH}
\author{Fran\c{c}ois~C.~Bocquet}\email{f.bocquet@fz-juelich.de (he/him/his)}	\affiliation{\pgi} \affiliation{\jara}

\date{\today $\quad$ --  \currfilename }

\begin{abstract}
  In the field of van der Waals heterostructures, the twist angle between stacked two-dimensional (2D) layers has been identified to be of utmost importance for the properties of the heterostructures. In this context, we previously reported the growth of a single layer of unconventionally oriented epitaxial graphene that forms in a surfactant atmosphere [F.~C.\ Bocquet \textit{et al.}, Phys.\ Rev.\ Lett.\ \textbf{125}, 106102 (2020)]. The resulting \Gnul layer is aligned with the SiC lattice, and hence represents an important milestone towards high quality twisted bilayer graphene (tBLG), a frequently investigated model system in this field.
  Here, we focus on the surface structures obtained in the same surfactant atmosphere, but at lower preparation temperatures at which a boron nitride template layer forms on SiC(0001). 
  In a comprehensive study based on complementary experimental and theoretical techniques, we find -- in contrast to the literature -- that this template layer is a hexagonal B$_x$N$_y$ layer, but \textit{not} high-quality hBN. It is aligned with the SiC lattice and gradually replaced by low-quality graphene in the 0$^\circ$ orientation of the B$_x$N$_y$ template layer upon annealing.
\end{abstract}
\maketitle

\section{Introduction}

The two-dimensional (2D) material graphene, as one of the next-generation materials for nanoelectronics, has attracted attention since Novoselov and Geim demonstrated its phenomenal electronic properties \cite{Novoselov2005, Geim2009}. It turned out that -- among other parameters that can be used to engineer its electronic properties -- the twist angle between the individual sheets in bilayer graphene stacks emerges as a promising parameter \cite{LopesDosSantos2007, Rozhkov2016}. Not only does the twisted bilayer graphene (tBLG) system show superconductivity at the magic angle of $1.1^\circ$ \cite{Cao2018_article, Cao2018_letter, Yankowitz2019, Yoo2019}, it is also expected to exhibit topological corner charges at $30^\circ$ twist angle, potentially making the \tB system a higher-order topological insulator \cite{Park2019}. So far, in most cases, the twist angle of bilayer graphene has been realized by stacking two micro-mechanically exfoliated graphene flakes under atmospheric or glove-box conditions \cite{Frisenda2018}. However, this method is neither scalable nor very well reproducible. For any type of large scale production, strategies involving epitaxial growth of graphene are desirable to circumvent these two disadvantages. Recently, we proposed a route to epitaxially grow high-quality single layer graphene that is well aligned with the SiC substrate (rotated $0^\circ$) \cite{Bocquet2020}. This material, named \Gnul in the following, represents the first (and decisive) step in preparing epitaxial \tB, since a conventionally oriented (rotated $30^\circ$) layer can be grown underneath. 

Hexagonal boron nitride (hBN) can also be exfoliated down to single atomic layers \cite{Britnell2012} and is stable at ambient conditions. Because of its large electronic band gap, it is used as an insulating material in hetero-stacks of 2D layers \cite{Geim2013, Wang2013}, as an insulating substrate \cite{Dean2010} and for encapsulating other 2D material layers and thus protecting them from contamination \cite{Ahn2016, Cadiz2017}. These applications reinforced the interest in growing hBN epitaxially. While hBN is frequently studied on metals (Refs.~\cite{Auwaerter2019, Felter2019, Raths2021} and references therein), only a few studies have been reported on semiconducting substrates \cite{Shimoyama2012, Shin2015}. Usually, the precursor molecule borazine (\borazine) is used for epitaxial growth of hBN. 
By annealing a SiC(0001) surface at sufficiently high temperatures ($1330$\C) in borazine atmosphere, high quality \Gnul can be produced at the wafer scale \cite{Bocquet2020}. When the same procedure is performed at lower temperatures, a boron nitride layer (\BNnul) is formed, 
having the same orientation as the \Gnul layer in the high-temperature case. This hexagonal \BNnul layer can be transformed into graphene at somewhat higher temperatures. While the orientation of the layer is conserved in this transition, the crystallinity of the thus formed graphene layer is not as good as that of the \Gnul layer produced directly at high temperatures in borazine atmosphere in Ref.\ \cite{Bocquet2020}. We hence address it as \lqGnul.

We determined the vertical and lateral structures of both the \BNnul and \lqGnul samples, using normal-incidence x-ray standing wave (NIXSW) and spot-profile analysis low energy electron diffraction (SPA-LEED). 
Angle-resolved photoelectron spectroscopy (ARPES) experiments revealed that, although LEED shows its hexagonal structure,  the \BNnul layer does not exhibit the electronic bands expected for hBN. This finding, as well as our NIXSW and density functional theory (DFT) data, show that it is \textit{not} a high-quality decoupled 2D hBN layer. Note that the precise stoichiometry of the \BNnul layer is also unknown, and that both the \BNnul and the \lqGnul layers are found to form atop a boron buffer layer (zeroth layer, ZL) at the interface to SiC.

\section{Methods}
\subsection{Sample preparation}\label{sec:preparation}
SiC samples were cut from a nitrogen-doped 6H-SiC(0001) wafer purchased from TankeBlue Semiconductor Co. Ltd. The surface was cleaned by direct current annealing for 30 minutes at 1050\C in ultra-high vacuum (UHV, pressure better than $1\times 10^{-9}$~mbar), while a Si flux was applied in order to compensate the sublimation of Si from the SiC surface \cite{Ramachandran1999}. The Si flux was produced by a heated Si wafer positioned $\approx 10$~cm above the sample surface.

For sample preparation, the SiC wafer was annealed for another 30 minutes at $880$\C, also under Si flux, in order to obtain the Si-rich \three reconstruction \cite{Riedl2007, Heinz2004}, which was confirmed using a multi channel plate (MCP)-LEED instrument. At a temperature below $880$\C, the Si flux was stopped and a borazine partial pressure of $1.5\times 10^{-6}$~mbar was applied. The sample temperature was then immediately increased to the desired value (between $1100$\C and $1250$\C) within less than five minutes and kept constant for 30 minutes while maintaining the borazine partial pressure. \blue{The \three reconstructed surface quickly transforms to a very reactive \sqthree \cite{Forbeaux1998} reconstruction, on which \BNnul layer forms. With this recipe, we minimize the time at which the \sqthree reconstruction is present, since it is well known to be very sensitive to impurities adsorbing from the residual gas \cite{Benesch2001}. This procedure is equivalent to annealing the \sqthree reconstructed surface in a borazine atmosphere, as performed in Ref.~\cite{Shin2015}, but expected to be less prone to contamination.} Afterwards, the samples were cooled down and transferred under UHV to the dedicated apparatus for the experiments. Depending on the annealing temperature, either a \BNnul or a \lqGnul layer forms on the surface, decoupled from the substrate by a boron ZL. Borazine was purchased from Katchem spol.\ s r.\ o., Praha, Czech Republic. 

As the preparation temperature was found to be a crucial parameter for the formation of the boron nitride and graphene layers, we developed a procedure to apply a specific temperature gradient along one lateral direction on the surface during borazine exposure. Local temperature measurements using a pyrometer revealed an approximately linear relation between the position on the sample (along the gradient direction) and the temperature. 
X-ray photoelectron spectroscopy (XPS) and LEED data could therefore be recorded with a temperature resolution (step width) of $\approx 10^\circ$C-$25^\circ$C. Small NIXSW data sets were also recorded for three different positions (temperatures) on these samples; however, the more extended and conclusive NIXSW data presented below were taken from samples prepared with a homogeneous temperature.

\subsection{ARPES and SPA-LEED}

All SPA-LEED and ARPES experiments were carried out at room temperature with an electron and photon beam footprint of approximately 3~mm$^2$. ARPES was performed at $40.8$~eV (He II) using a Scienta R4000 hemispherical electron analyzer ($28^\circ$ electron acceptance angle) and a Scienta VUV5k monochromatized Helium source. SPA-LEED images were recorded with an Omicron SPA-LEED instrument, which has a transfer width $>1000$~\AA, corresponding to a $k$-space resolution better than $0.006$~\AA$^{-1}$ or $0.25~\%$BZ$_\textrm{SiC}$. The images shown in this work are distortion-corrected using the LEEDLab / LEEDCal software v.~1.1. \cite{LEEDLab, Sojka2013}.

\subsection{XPS and NIXSW}

The NIXSW technique \cite{Zegenhagen2013, Zegenhagen1993, Woodruff1998, Woodruff2005} is a  model-free method to probe vertical distances, in our case, between the overlayer and the topmost atoms of the SiC bulk. It comes with chemical sensitivity since it is based on XPS, but these data have to be recorded at relatively high photon energies (hard x-rays, see below). Often, XPS data are additionally recorded using soft x-rays, offering a better energy resolution. This allows one to unambiguously identify the spectral features in the XPS data and to set up a fitting model for the hard x-ray data.

Both types of experiments were carried out at the beamline I09 of the Diamond Light Source Ltd., Didcot, UK. The beamline provides soft and hard x-ray beams, both focused on the same sample position with a footprint of approximately $400 \times 250~\mu$m$^2$ at normal incidence to the surface. Photoelectrons are detected by a VG Scienta EW4000 HAXPES hemispherical electron analyzer with an acceptance angle of $56^\circ$. All XPS and NIXSW data presented in this work were measured in normal incidence and grazing emission geometry, collecting photoelectrons with emission angles between $\phi=62^\circ$ and $90^\circ$ with respect to the surface normal.

For the NIXSW measurements, the sample is aligned with the x-ray beam such that the Bragg condition for a chosen reflection $\textbf{H}=(hkl)$ of the bulk crystal is fulfilled close to normal incidence of the x-rays with respect to the Bragg planes. This condition defines the (hard x-ray) photon energy that has to be used for the NIXSW experiment. Then, core-level spectra for all relevant species are recorded simultaneously with the Bragg-diffracted x-ray intensity, while the photon energy $h\nu$ is scanned through the Bragg condition. 
During such a photon energy scan, the phase of the standing wave changes from $\pi$ to $0$, causing the standing wave to traverse half of the Bragg plane spacing $d_{(hkl)}$. As a  consequence, for an atom at a specific height $z$, the yield of the emitted photoelectrons is modulated by the shifting x-ray standing wave field. By integrating the individual spectra and plotting their intensity vs.\ the photon energy, one thus obtains an NIXSW yield curve $Y(h\nu)$ that is characteristic for the (average) $z$-position of that atomic species. 

From fitting the photoelectron yield curves one obtains two structural parameters: the coherent position \Pc and the coherent fraction \Fc, both ranging from zero to one. \Pc represents the height of the probed atomic species above the next Bragg plane below, in units of $d_{(hkl)}$. 
\Fc is a measure of the vertical order (``vertical'' in the sense of ``perpendicular to the (hkl) Bragg planes''). \mbox{\Fc = 1} indicates that all atoms of the specific species are located at the same height above the next Bragg plane below, and values of \Fc significantly below $1.0$ indicate some (vertical) disorder or multiple adsorption heights. 

In our case, we performed the NIXSW measurements with the (0006) reflection of a 6H-SiC bulk crystal, having a Bragg plane spacing of $d_{(0006)}=2.520$~{\AA}. We recorded full data sets for the C~$1s$, Si~$2s$, B~$1s$ and N~$1s$ core levels. The analysis was performed using the software package \textsc{Torricelli} \cite{Bocquet2019, Torricelli}. The influence of nondipolar effects and the deviation from normal incidence geometry (given by the experimental conditions, namely a fixed Bragg angle of $86.5^\circ$) were taken into account, as well as the large acceptance angle of the electron analyzer. For more details see Ref.~\cite{vanStraaten2018}.

\subsection{Density functional theory}\label{sec:dft}

We performed DFT calculations using the all-electron electronic structure code FHI-aims \cite{fhiaims_blum_2009} and the ELSI infrastructure for large-scale calculations \cite{Yu2018, Yu2020}. The default numerical settings “light” of FHI-aims and the Perdew-Burke-Ernzerhof (PBE) exchange-correlation functional \cite{pbe_perdew_1996} with the Tkatchenko-Scheffler van der Waals correction \cite{ts_tkatchenko_2009} were employed.
This level of theory has proven to accurately describe the structure of epitaxial graphene on SiC~\cite{sic_nemec_2013,sic_sforzini_2015,sic_tu_2016}. A $5 \times 5 \times 1 $ slab model of 6H-SiC was employed, with the bottom-most carbon atoms terminated by hydrogen atoms to mimic the bulk material used in experiments. A boron ZL and a hBN monolayer were placed on top of the SiC substrate, with the initial interlayer distances matching the values obtained by NIXSW. As the structure of the ZL is unknown, we considered a series of randomly-generated boron monolayer models with a well-defined number of boron atoms. The atoms in each model were relaxed until the maximum force acting on the atoms was below 0.01 eV/\AA.

\section{Results and Discussion}
In Sec.~\ref{sec:temp-gradient}, we present XPS and LEED measurements performed on temperature gradient samples, i.e., with a rather good temperature resolution. These gradient samples allowed us to study the effect of the preparation temperature in detail.
On selected samples, which were prepared with homogeneous preparation temperatures, we investigated the lateral structure in more detail using SPA-LEED (Sec.~\ref{sec:SPA}), the vertical structure using NIXSW (Sec.~\ref{sec:NIXSW}), and the electronic structure using ARPES (Sec.~\ref{sec:electronic+air}). The comprehensive analysis based on these complementary techniques, together with DFT calculations performed for hBN/SiC(0001) (Sec.~\ref{sec:DFT}), allows us to study the transformation from the \BNnul layer to the \lqGnul layer taking place in the preparation temperature range from 1150\C to 1250\C.

\subsection{Preparation temperature dependency of the layer structure and the chemical composition}\label{sec:temp-gradient}

\begin{figure}
	\includegraphics[width=\linewidth]{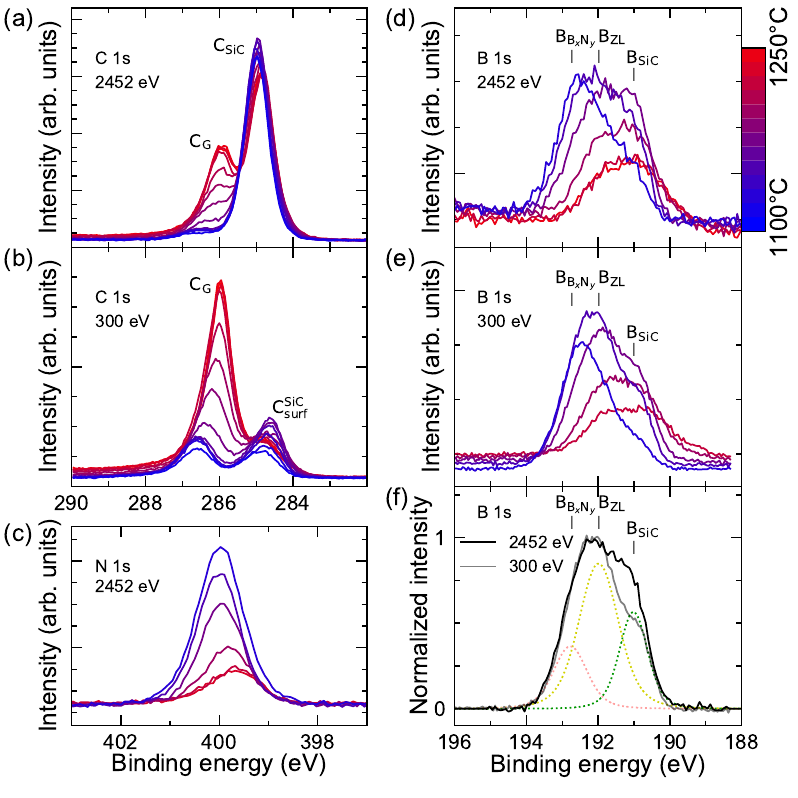}
	\caption{\label{fig:XPSevolution} 
		C~$1s$ (a,b), N~$1s$ (c), and B~$1s$ (d-f) core-level spectra measured with hard and soft x-rays on a temperature gradient sample at different positions corresponding to the preparation temperatures color-coded from blue (low T) to red (high T). A comparison of hard (a and d) with soft x-rays (b and e) allows us to identify surface and bulk species for C and B, see text. In (f), this is demonstrated for a B~$1s$ spectrum recorded with higher statistics on the \BNnul sample. The data are shown after background subtraction, normalization and calibration of the binding energy scale. All spectra have been measured in a grazing emission geometry. }
\end{figure}

In Fig.\ \ref{fig:XPSevolution}, core-level spectra are shown, obtained using both hard and soft x-rays. The data were recorded at different positions on the temperature gradient sample, corresponding to preparation temperatures between 1100\C and 1250\C, as color-coded from blue to red. We were able to identify two C~1$s$ and three B~1$s$ components. The fact that hard and soft x-rays have different probing depths allows us to find out where the individual components stem from. For C~1$s$, the situation is clear: The component at $284.8$~eV is a bulk species (\CSiC) since it is stronger at higher photon energy and almost independent of preparation temperature, see Fig.\ \ref{fig:XPSevolution}(a) and (b). The peak at $286.0$~eV is a surface component, very dominant at the smaller photon energy, and only present for higher preparation temperatures. It hence stems from the \lqGnul layer (\CGnul). The N~1$s$ spectra in Fig.\ \ref{fig:XPSevolution}(c) show only one peak, the intensity of which is dropping and shifting to lower binding energy with increasing temperature. For B~1$s$, it is more difficult to identify the components. Figure~\ref{fig:XPSevolution}(f) reveals a bulk-like behavior for the component at $190.8$~eV (\BSiC). Figures~\ref{fig:XPSevolution}(d) and (e) also show that this component is present at all preparation temperatures. Hence, some of the boron atoms must have diffused into the bulk, an effect that has already been reported earlier \cite{Sforzini2016b}. The other two components at $192.6$~eV and $191.8$~eV (labeled \BBN and \BZL) are located closer to the surface and stem from the \BN layer and from the boron ZL underneath, respectively. 

\begin{figure}
  \includegraphics[width=\linewidth]{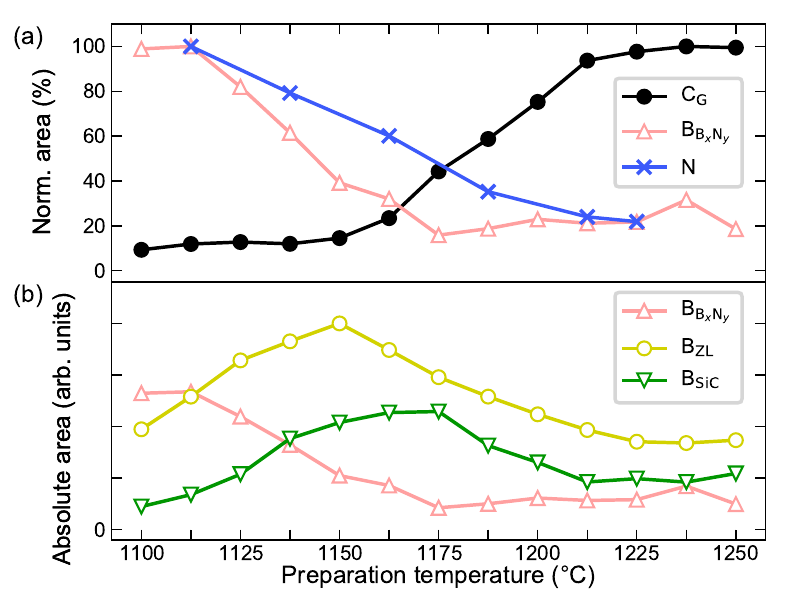}
  \caption{\label{fig:Gradient-summary} Integrated components obtained from the core-level spectra shown in Fig.\ \ref{fig:XPSevolution} (a), (c) and (d) as a function of the preparation temperature. (a) \blue{\CGnul, \NBN, and \BBN intensities normalized to their respective maximum.} (b) Absolute intensities of the three B~$1s$ components. }
\end{figure}

Figures \ref{fig:XPSevolution}(c) and (d) reveal that the N~1$s$ and \BBN components decrease in their intensities with increasing preparation temperature in a similar way. This evolution is better quantified in Fig.\ \ref{fig:Gradient-summary}(a), showing normalized intensities of the core-level spectra. It suggests that both the \BBN and N~1$s$ components stem from the same boron nitride structure, which disappears with increasing preparation temperature. At the same time, the \CGnul component increases, indicating that the \lqGnul layer is formed as the \BN layer disappears. We note  that the ratio $x/y$ in the \BN layer appears to have a small preparation temperature dependency (Fig.\ \ref{fig:Gradient-summary}(a)).

In Fig.\ \ref{fig:Gradient-summary}(b), the integrated intensities of the three B~$1s$ components are shown as a function of the preparation temperature. For $\approx 1100$\C, \BBN is the dominant component, the \BZL component is also clearly visible, while the \BSiC component is small. With \BBN disappearing, the other two components increase, most likely just indicating that the B-N bonding in the \BN layer is being broken, which causes a core-level shift towards smaller binding energies. Both the \BZL and the \BSiC curves are running through a maximum at $1150^\circ\text{C}-1175$\C. At the end of the preparation temperature scale ($1250$\C), the \BZL component has basically the same intensity as in the beginning (at $1100$\C), while \BSiC is slightly more intense. This indicates that the boron atoms from the \BN layer in the end either evaporate or diffuse deeper into the bulk, so that they are not detected any more. 

The scenario to be deduced from these XPS measurements is straightforward: At a preparation temperature above $\sim 1150$\C the \BN layer, which is located above a boron ZL (\BZL), is gradually replaced by the \lqGnul layer (\NBN and \BBN decrease and \CGnul increases), while neither the bulk nor the boron ZL are largely affected.

\begin{figure}[t]
  \includegraphics[width=\linewidth]{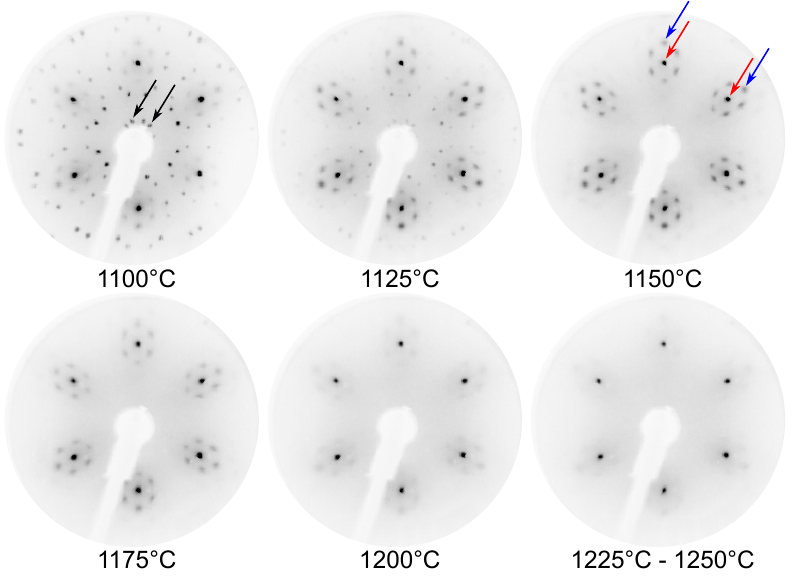}
  \caption{\label{fig:JaneLEED} Series of LEED images recorded at different positions on the temperature gradient sample, corresponding to preparation temperatures between 1100\C and 1250\C. Blue, red and black arrows indicate diffraction spots of \BNnul, SiC, and the \extraspots superstructure, respectively. The electron energy was $100$~eV.}
\end{figure}

\begin{figure*}
  \begin{center}
    \includegraphics[width=0.9\textwidth]{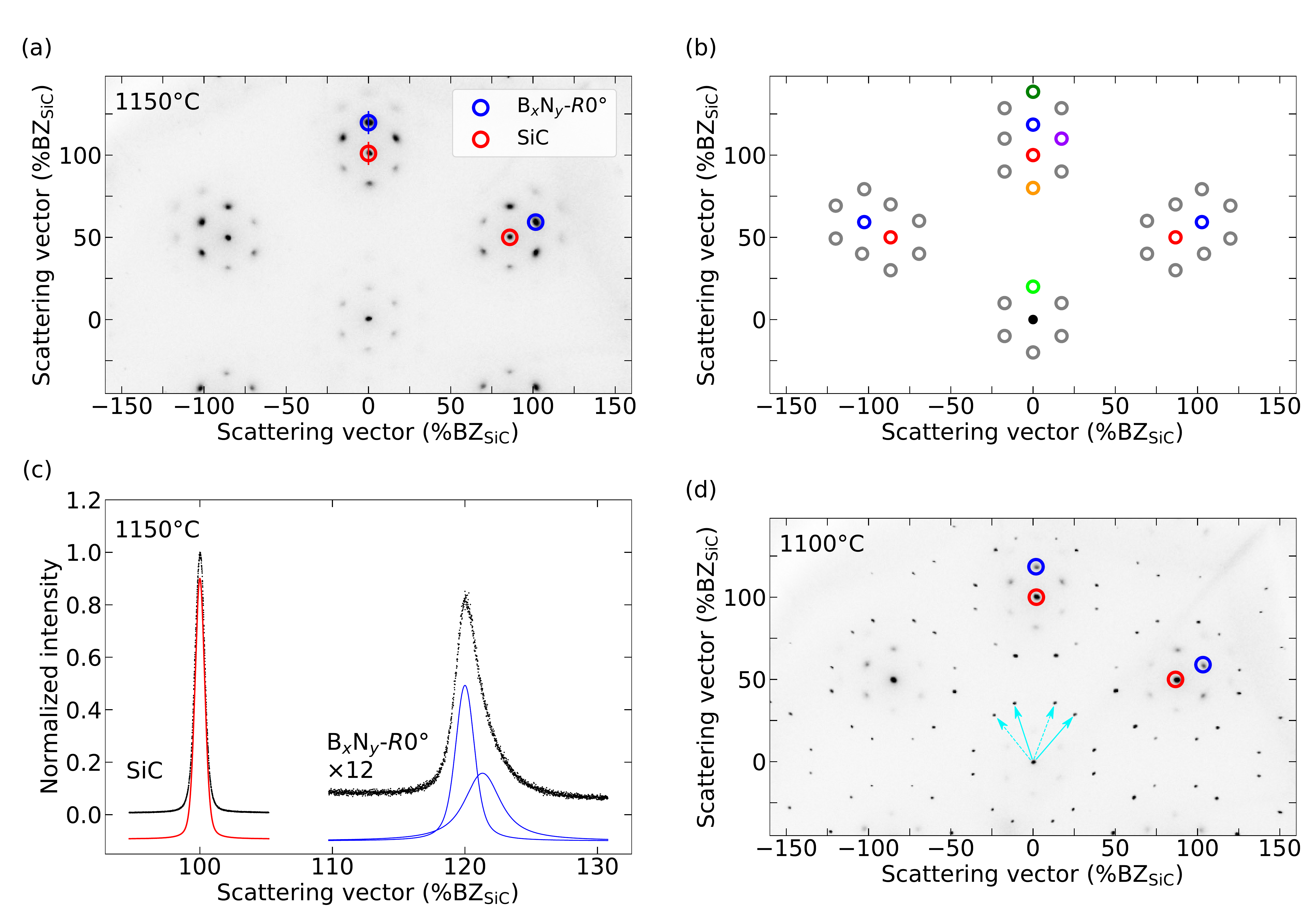}
    \caption{\label{fig:SPALEED}
      (a) Distortion-corrected SPA-LEED pattern of a homogeneous \BNnul sample prepared at 1150\C. The (10) and (01) LEED spots of the SiC bulk and the \BNnul layer are marked by red and blue circles, respectively. The short blue and red lines through the (01) spot of the \BNnul layer and SiC indicate where the radial line scans shown in (c) were recorded. The electron energy was $165$~eV. (b) Illustration of the LEED spot positions. Red and blue circles are marked as in (a). All other spots originate from double diffraction processes involving both the SiC bulk and the \BNnul layer, colors are explained in the text. (c) Radial line scans through the \blue{SiC and} \BNnul (01) diffraction spots. \blue{The SiC peak is fitted with a single symmetric Voigt profile. The \BNnul peak is fitted with two symmetric Voigt profiles.} (d) As (a), but for a \BNnul sample prepared at 1100\C. } 
  \end{center}
\end{figure*}

For the same gradient sample we have also recorded LEED patterns that can be correlated to the core-level spectra. Figure \ref{fig:JaneLEED} shows six diffraction patterns, corresponding to different preparation temperatures. 
The SiC(0001) (10) and (01) spots are marked with red arrows in the pattern for $1150$\C.
For preparation temperatures between $1125$\C and $1175$\C, the LEED patterns are dominated by groups of six reflections forming a hexagon around each of the SiC first order bulk spots. Their intensities reach a maximum at $1150$\C, decrease afterwards, and become very small at $1200$\C and above. As demonstrated in the following (Sec.~\ref{sec:SPA}), these spots can be attributed to the \BN layer.
The outermost spots of each of the hexagons (those marked with blue arrows)
are the \BN (10) and (01) spots, all others are explained by multiple diffraction effects.
Note that the azimuthal orientation of the \BN LEED pattern relative to that of the bulk clearly indicates that the \BN layer is aligned with the substrate lattice ($0^\circ$ rotation, \BNnul).
For $1200$\C and above, the $\{10\}$ and $\{01\}$ spots become very weak and move slightly towards larger $k_{||}$. This indicates the transformation from the \BNnul layer to the \lqGnul layer, driven by carbon atoms replacing boron and nitrogen \cite{Bocquet2020}.

Note that the \lqGnul layer can be obtained by ramping the SiC(0001) temperature to $1225$\C, as described in the methods section (sec.\ \ref{sec:preparation}), or by post-annealing the readily prepared \BNnul layer in UHV, as discussed in the supplement of Ref.\ \cite{Bocquet2020}. However, the high quality of the \Gnul layer that is obtained when exposing the SiC sample to borazine directly at higher temperatures ($1330$\C, see Ref.\ \cite{Bocquet2020}) cannot be reached by post-annealing, neither of the \BNnul nor the \lqGnul samples.

Finally we mention that in the 1100\C LEED pattern, additional sharp spots of a \extraspots superstructure are visible (black arrows in Fig.\ \ref{fig:JaneLEED}). They vanish already below 1150\C, i.e., before the \NBN and \BBN components disappear, indicating that the underlying reconstruction is lifted even before the \BNnul $\rightarrow$ \lqGnul layer transformation process takes place (see below). 

\subsection{Lateral structure from SPA-LEED}\label{sec:SPA}

The data presented so far were recorded from temperature gradient samples, illustrating the effect of the preparation temperature on the layer formation.
In the following, we show results obtained from detailed investigations on homogeneous samples prepared at different temperatures. At first we analyze the LEED patterns in detail.

Figure~\ref{fig:SPALEED}(a) displays a SPA-LEED pattern of a sample prepared at $1150$\C. The (10) and (01) spots of the SiC bulk are marked by red circles, those of the \BNnul layer by blue circles. These are the only spots visible in the LEED image that are due to single diffraction. All other spots are due to multiple diffraction processes of the \BNnul layer and the SiC substrate, as illustrated in Fig.\ \ref{fig:SPALEED}(b): The satellites around the (00) spot are double-diffraction spots involving one first order spot each of the \BNnul layer and the SiC substrate. For example, the one indicated by a light green circle is due to a double-diffraction process of the $(01)_\textrm{BN}$ and the $(0\overline{1})_\textrm{SiC}$ reflection.
Those located around the SiC first order reflections involve some higher order \BNnul or SiC diffraction spots, e.g., $(02)_\textrm{BN} + (0\overline{1})_\textrm{SiC}$ (dark green), $(0\overline{1})_\textrm{BN} + (02)_\textrm{SiC}$ (orange), and $(10)_\textrm{BN} + (\overline{1}1)_\textrm{SiC}$ (magenta). Spots marked by gray circles can be explained in a similar way. Note that all spots involving a \textit{first} order \BNnul reflection are relatively strong and form the hexagon around the first order SiC reflections. Spots involving the second order \BNnul reflections are weaker and not always visible. The disappearance of the \BNnul $\{10\}$ and $\{01\}$ reflections and all double diffraction spots at preparation temperatures between $1175$\C and $1200$\C (see Fig.\ \ref{fig:JaneLEED} and discussion above) indicates the transformation of the \BNnul layer to the \lqGnul layer in this temperature range. 

Owing to the high $k$-space resolution of the SPA-LEED technique, radial line scans through the \BNnul spots reveal an asymmetric profile, \blue{in contrast to the SiC bulk spots,} see Fig.\ \ref{fig:SPALEED}(c). This was found for both the \BNnul first order (single diffraction) reflections and all double diffraction spots involving a BN reflection. \blue{The peak broadening of these spots is symmetric with respect to SiC first order reflections. The satellites are always broadened on the side facing away from the SiC bulk peak, confirming that} the satellites are double diffraction peaks involving the BN first order reflections. The asymmetric peak can be fitted by two symmetric Voigt profiles with a distance in k-space of about $2\%$BZ$_\mathrm{SiC}$, see Fig.\ \ref{fig:SPALEED}(c). While the main peak clearly stems from the \BNnul layer, the side peak is possibly the first order Bragg reflection of the \lqGnul layer. This is indicated by the difference in lattice parameters between the \BNnul and \lqGnul layers, which matches the separation of the two peaks. The \lqGnul Bragg peak is quite weak, owing to the very early stage of the \lqGnul layer formation at this temperature. This is in agreement with the small \CGnul component in the C~$1s$ XPS spectrum at the lower end of the preparation temperature scale, see Fig.\ \ref{fig:XPSevolution}(a) and (b).

Using the LEEDLab software \cite{Sojka2013, LEEDLab}, we have corrected the SPA-LEED images for distortions and fitted the spot positions in order to determine the lattice parameters of the involved structures. We find lattice parameters of $3.08(4)$~\AA\ for the SiC bulk and $2.60(3)$~\AA\ for the \BNnul layer. 
The latter indicates a 
$(3.6\pm1.2)$\% expansion compared to literature values for hBN ($2.51$~{\AA} \cite{Ooi2006}), i.e., the \BNnul layer is significantly less densely packed than a 2D hBN layer. 
From the width of the \BNnul SPA-LEED peaks, in comparison to that of the SiC bulk, we estimate the average domain size within the \BNnul layer. The main component of the line scan shown in Fig.\ \ref{fig:SPALEED}(c) has a full width at half maximum (FWHM) of $w=1.709(8)$\BZ, which is approximately twice the width of the SiC bulk peaks ($w=0.853(1)$\BZ). From these numbers we estimate \blue{a lower limit to the average domain size of $2\pi/w=30$~nm} for the \BNnul layer.

\begin{figure*}[!ht]
  \includegraphics[width=\textwidth]{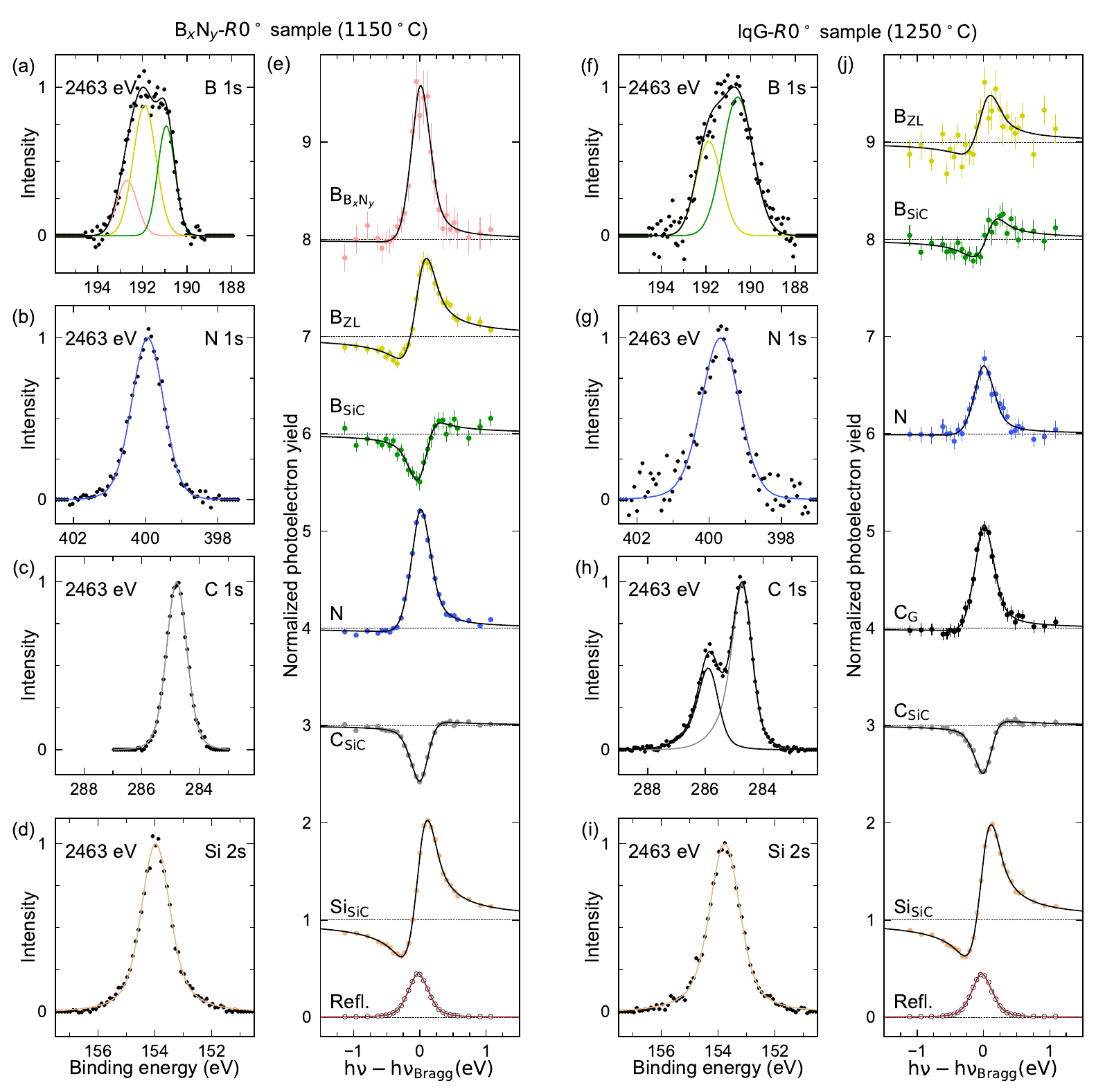}
  \caption{\label{fig:yieldcurves} 
    NIXSW results for the \BNnul sample prepared at 1150\C and the \lqGnul sample prepared at 1250\C. (a-d, f-i) Core-level spectra recorded at a photon energy $\approx 2$~eV below the (0006) Bragg energy. Single peaks (solid lines) were fitted to the data, except for the B~$1s$ (both samples) and C~$1s$ spectra (\lqGnul sample only), which were fitted using two- or three-component models. (e, j) Yield curves of the probed core-levels and reflectivity curve of the SiC(0006) Bragg reflection. In (e), the yield curves for \BBN, \BZL, \BSiC, \NBN, and \CSiC are displaced vertically by 7, 6, 5, 3, and 2, in (j) those for \BZL, \BSiC, \NBN, \CGnul and \CSiC by 8, 7, 5, 3, and 2, respectively.}
\end{figure*}

In Fig.\ \ref{fig:SPALEED}(d) we show a SPA-LEED pattern of a sample prepared at a lower preparation temperature ($1100$\C). Beside the \BNnul and SiC diffraction spots, this pattern also shows additional spots stemming from a \extraspots superstructure. The unit cell is indicated by cyan arrows (solid and dashed for two mirror domains) in Fig.\ \ref{fig:SPALEED}(d). As mentioned above, this pattern disappears quickly when higher preparation temperatures are applied, clearly before the \BBN and \NBN core-level components vanish.
Furthermore, the reflections are sharp and rather intense in relation to the \BNnul \{10\} and \{01\} reflections, suggesting that they do \textit{not} originate from the \BNnul layer, but rather from a boron-induced reconstruction of the SiC(0001) surface (we will see below that a \BZL layer is also present at the interface to SiC at this preparation temperature).
Based on the preparation temperature ($1100$\C), we suggest that this reconstruction consists of B and Si adatoms, since it is known that the Si-rich SiC(0001)-($3\times3$) surface is stable in UHV up to $\approx 1050$\C \cite{Forbeaux1998}. The \extraspots surface reconstruction is destroyed before the \BNnul $\rightarrow$ \lqGnul layer transformation, i.e., already at preparation temperatures slightly larger than $1100$\C.

\subsection{Vertical structure from NIXSW}\label{sec:NIXSW}

The vertical structure of the \BNnul and \lqGnul samples was determined by NIXSW. Typical core-level spectra and the yield curves extracted from the XPS data are shown in Fig.\ \ref{fig:yieldcurves}(a)-(e) for the \BNnul sample. The preparation temperature for this sample was $1150$\C. Data corresponding to a lower preparation temperature ($1100$\C) were also recorded, but in a much smaller data set since it was taken from a temperature gradient sample. For both temperatures very similar results were obtained, see below. In Fig.\ \ref{fig:yieldcurves}(f)-(j), we show the same type of data from a \lqGnul sample prepared at $1250$\C.

The first crucial step in the analysis of NIXSW data are finding the best fitting model for the XPS data. For the N~$1s$ and Si~$2s$ species this is straightforward, since the spectra contain only one slightly asymmetric peak and can be fitted well with one (asymmetric) Voigt profile (Fig.\ \ref{fig:yieldcurves}(b), (d), (g), and (i)). Also the C~$1s$ data (Fig.\ \ref{fig:yieldcurves}(c), (h)) are easily fitted, since the peaks stemming from SiC and \lqGnul are well separated (see also Sec.~\ref{sec:temp-gradient}). 

For B~$1s$ the analysis is more difficult. As already discussed in Sec.~\ref{sec:temp-gradient}, there are three components showing relatively small core-level shifts with respect to each other. We find that the data set recorded on the \BNnul sample can be best fitted with all three components under tight constraints, namely fixed binding energy differences of the \BBN and \BZL components relative to the \BSiC peak ($1.76$~eV and $0.97$~eV, respectively). The B~1$s$ fitting model is shown in Fig.\ \ref{fig:XPSevolution}(f) and Fig.\ \ref{fig:yieldcurves}(a). 

\begin{figure*}
  \includegraphics[width=0.7\linewidth]{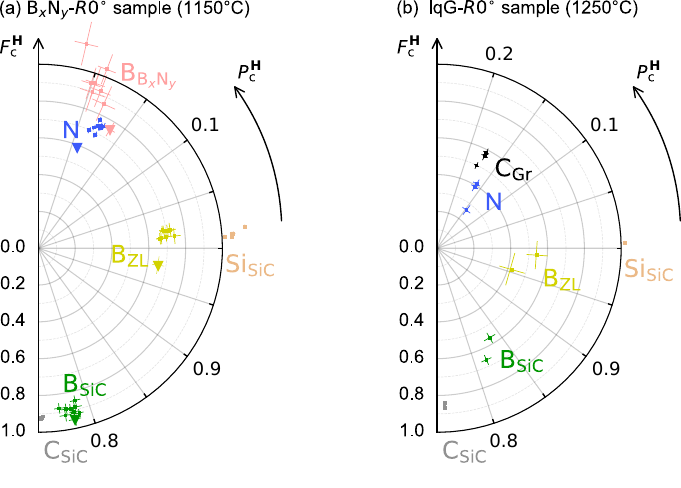}\\
  \caption{\label{fig:NIXSW_Argand}
    Argand diagrams summarizing the NIXSW fit results for the (a) \BNnul and (b) \lqGnul samples on SiC. The results of all individual measurements are shown as small squares that represent a polar vector with \Fc as its length and \Pc as its polar angle. In (a), we also show the B~1s and N~1s results corresponding to a preparation temperature of 1100\C (large triangles), as obtained from a temperature gradient sample.}
\end{figure*}

The models in Fig.\ \ref{fig:yieldcurves}(a)-(d) and Fig.\ \ref{fig:yieldcurves}(f)-(i) were used to extract NIXSW yield curves from the XPS data. We recorded several yield curves at different spots on the surface, and show a representative selection together with a typical reflectivity curve of the SiC(0006) reflection in Fig.\ \ref{fig:yieldcurves}(e) and (j). From fitting the yield curves, the coherent fractions \Fc and coherent positions \Pc are obtained, and their averages listed in Table~\ref{table:results}. In Fig.\ \ref{fig:NIXSW_Argand}(a) we show an Argand diagram which illustrates the results of all individual scans, as vectors in a polar diagram. The data points represent the heads of vectors with \Fc as their length and \Pc as their polar angle. We show results obtained from a large data set recorded for a homogeneous sample prepared at 1150\C (all individual scans shown as small squares), and also for a small data set recorded from a temperature gradient sample at 1100\C (larger triangles). The results for these two preparation temperatures are very similar. 

The Argand diagram in Fig.\ \ref{fig:NIXSW_Argand}(a) illustrates very well the basic findings of our NIXSW analysis: Most coherent fractions are sufficiently high to indicate single-height adsorption of the individual species, with some slight disorder in some cases, see below. In particular, the three distinct boron species are clearly confirmed, since their coherent positions are very different. 

For both \BNnul samples, \BBN and \NBN have very similar coherent positions \Pc, indicating that these species are located within one layer with only a small buckling. The \BZL layer below is similarly flat, since its coherent fraction is close to that of \NBN in the \BNnul layer, although smaller than that of \BBN. The third boron species, \BSiC, is attributed to boron atoms diffusing into the bulk. Its coherent fraction is high, indicating that boron atoms adopt well-defined doping sites which are almost precisely located in the carbon layers of the SiC bulk, since the coherent positions for \BSiC and \CSiC are almost identical (see Table \ref{table:results})).

\begin{table}[b]
	\caption{\label{table:results} NIXSW results (averaged values from measurements on several spots on the sample) for the \BNnul sample prepared at 1150\C and the \lqGnul sample prepared at 1250\C. The distances are given with respect to the topmost Si atoms of the substrate: For a species $X$, the distance was calculated as $z_X=(N_X+P_X^\mathbf{H}-P_\mathrm{Si_{SiC}}^\mathbf{H})\times d_{(hkl)}$, with the number of Bragg planes $N_X$ located between the atomic species and the substrate surface plane, and the Bragg layer spacing $d_{(hkl)} = d_{(0006)} = 2.520$~{\AA}. For \BSiC the $z$ position was not calculated since this species diffuses into the bulk.}
	\renewcommand{\arraystretch}{1.1}
	\begin{tabular}{p{0.6cm}p{0.3cm}p{1.05cm}p{1.1cm}p{1.15cm}p{1.15cm}p{1.1cm}p{1.15cm}}
		\hline 
		&  &\multicolumn{3}{c}{\BNnul sample} & \multicolumn{3}{c}{\lqGnul sample} \\
		$X$ &$N_X$& ~~~\Pc  & ~~\Fc   & ~$z_X$ [\AA] & ~~~\Pc  & ~~\Fc  & ~$z_X$ [\AA] \\
		\hline
		B$_{\mathrm{B}\!_x\!\mathrm{N}\!_y}$   
		& ~2 & ~0.19(1) & 0.95(10)& 5.50(3) &     ~~~~--   & ~~~--    & ~~~--  \\
		\NBN   & ~2 & ~0.18(1) & 0.73(2) & 5.46(3) &     ~0.16(2) & 0.36(7)  & 5.43(5)\\
		\CGnul & ~2 & ~~~~--   & ~~~--   & ~~~--   &     ~0.18(1) & 0.55(4)  & 5.47(3)\\
		\BZL   & ~1 & ~0.02(1) & 0.69(3) & 2.54(3) &     -0.02(2) & 0.50(10) & 2.45(5)\\
		\SiSiC & ~0 & ~0.01(1) & 1.06(4) & 0.0     &     ~0.00(1) & 1.02(1)  & 0.0    \\		
		\BSiC  &    & ~0.78(1) & 0.90(3) & ~~~--   &     ~0.82(1) & 0.62(8)  & ~~~--  \\		
		\CSiC  & -1 & ~0.75(1) & 0.92(1) &-0.65(3) &     ~0.76(1) & 0.85(1)  &-0.62(3)\\
		\hline 
	\end{tabular}
\end{table}

In the sample prepared at 1250\C, the \BNnul layer has given way to a \lqGnul layer, as discussed in Sec.~\Ref{sec:temp-gradient}. Hence, when fitting the B~$1s$ spectra, the best results were obtained without the \BBN component in the model, and with the width of the \BSiC component constrained to $1.2$ times of that of \BZL (see Fig.\ \ref{fig:yieldcurves}(f)). Alternatively, we have also tried to use the three-component fitting model used for the \BNnul sample, but the results were less reliable. For the other spectra (C~$1s$, N~$1s$ and Si~$2s$) we applied the same model as the one used in the data analysis for the \BNnul sample. 

\begin{figure*}
	\includegraphics[width=\linewidth]{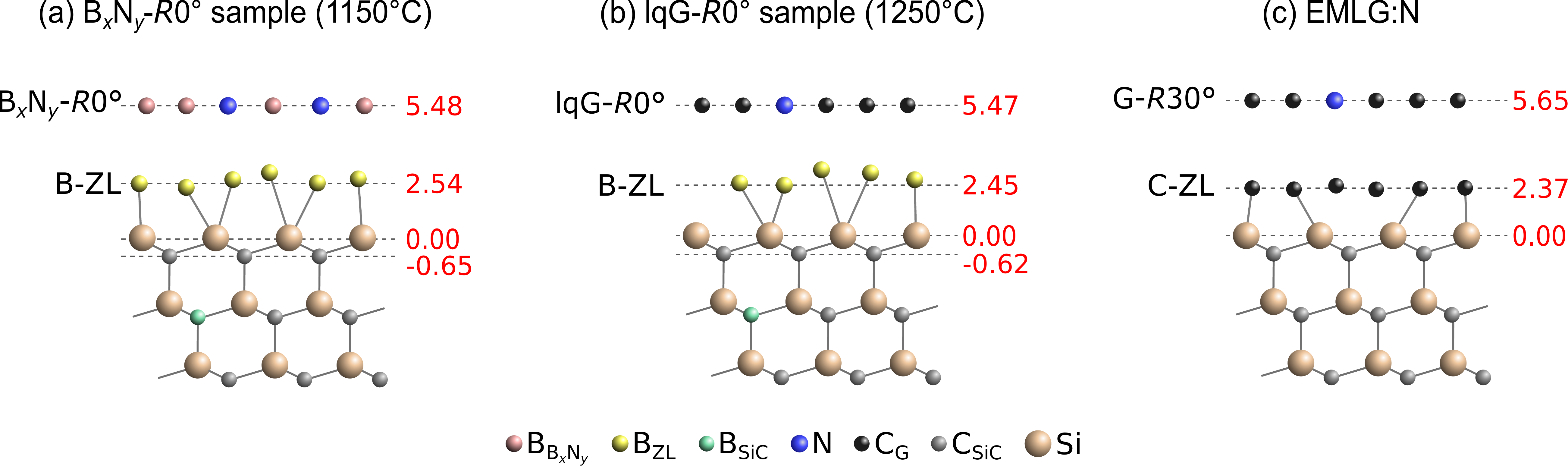}\\
	\caption{\label{fig:structure} 
		Ball and stick models of the vertical structures of (a) \BNnul, (b) \lqGnul and (c) EMLG:N on SiC. The latter is reproduced from Sforzini et al.\ \cite{Sforzini2016}. Note that in (a) and (b) the ZL consists of boron, in (c) of carbon. Numbers represent vertical distances to the uppermost Si layer in \AA ngstrom. }
\end{figure*}

The NIXSW yield curves obtained for the \lqGnul sample are presented in Fig.\ \ref{fig:yieldcurves}(j). The NIXSW results are shown in the Argand diagram in Fig.\ \ref{fig:NIXSW_Argand}(b) and their averages are listed in Table~\ref{table:results}. It is obvious that some coherent fractions are smaller than those of the \BNnul sample, in particular the ones for \BZL, \BSiC and \NBN. For the latter this is easy to understand, since the \BNnul layer does not exist any more and the small number of remaining N atoms occupy a less well-defined vertical position than before. Its coherent position, however, which is very close to that of carbon in the \lqGnul layer, indicates that the majority of the remaining nitrogen atoms incorporates as a dopant of the \lqGnul layer, similar to Ref.~\cite{Sforzini2016}. For the two boron species \BZL and \BSiC, the low coherent fractions reveal that more vertical disorder is introduced by the higher preparation temperature. This is in particular interesting for the boron ZL, since it might explain why the \lqGnul layer above also exhibits a significantly smaller coherent fraction compared to the \BNnul layer.
The coherent fraction of the species in the uppermost layer (0.73 for \NBN and 0.95 for \BBN) reduce to 0.55 for \CGnul in \lqGnul. This is consistent with the fact that the \lqGnul layer is of lower quality compared to the sample produced by the alternative scenario at even higher preparation temperatures, where no \BNnul layer is formed \cite{Bocquet2020}. It also agrees with the finding of a broad and faint Dirac cone in the ARPES experiments (see below) and weak and blurry LEED spots (Fig.\ \ref{fig:JaneLEED}). 

\begin{table}[b]
	\caption{\label{table:BondingDistance} Analysis of bonding distances. The distances between the layers (\BNnul layer to ZL, \lqGnul layer to ZL, and ZL to substrate) as obtained from NIXSW, are listed in \AA ngstrom and in percent of the corresponding van der Waals bonding distance (vdW). For the \BNnul and \lqGnul samples, the ZL consists of boron, for the \blue{EMLG:N sample}, the ZL consists of carbon. Van der Waals radii are taken from Ref.\ \cite{Mantina2013}.}
	\begin{tabular}{p{1.55cm}p{0.75cm}p{0.75cm}p{1.0cm}p{0.8cm}p{1.0cm}p{0.8cm}p{0.9cm}}
	\hline
	& vdW    & \multicolumn{6}{c}{Distances [\AA] and [\%vdW] } \\
	Bond type             & ~[\AA] & \multicolumn{2}{c}{\BNnul} & \multicolumn{2}{c}{\lqGnul} & \multicolumn{2}{c}{EMLG:N \cite{Sforzini2016}}  \\
	\hline
	\BBN - ZL               & 3.84 & ~2.96 & 77.1\% & ~ & ~ & ~ & ~ \\
	~ ~ ~ \NBN ~- ZL        & 3.47 & ~2.92 & 84.1\% & ~2.98 & 85.9\% & ~3.35 & 96.5\% \\
	~ C$_\textrm{lqG}$ - ZL & 3.62 &       &        & ~3.02 & 83.4\% & ~3.28 & 90.6\% \\
	~ ~ ZL\ - Si            & 4.02 & ~2.54 & 63.2\% & ~2.45 & 60.9\% & ~2.37 & 59.0\% \\
	\hline
    \end{tabular}

\end{table}

Based on these results, we present structural models for the \BNnul and the N-doped \lqGnul samples as shown in Fig.\ \ref{fig:structure}. It is remarkable that the \BNnul layer in the \BNnul sample and the \lqGnul layer in the \lqGnul sample are found to be at almost the same height above the bulk surface -- $5.48$~{\AA} (average of the B and N heights) and $5.47$~{\AA}, respectively -- while the boron ZL is (in absolute numbers) slightly closer to the substrate for the \lqGnul sample ($2.45$~{\AA}) compared to the \BNnul sample ($2.54$~\AA). However, in units of the expected van der Waals (vdW) bonding distances, the \BNnul to ZL distance (2.94~\AA) and the \lqGnul to ZL distance (3.02~\AA) are almost identical: $80.6$\% for the average of B and N in the \BNnul layer, and $83.4$\% for C in the \lqGnul layer, see Table \ref{table:BondingDistance}.

A comparison to values obtained by Sforzini \textsl{et al.}\ \cite{Sforzini2016} for the N-doped \blue{epitaxial monolayer graphene (EMLG:N) sample}, which rests on a \textit{carbon} ZL (graphene buffer layer), is instructive: The atomic model is shown in Fig.\ \ref{fig:structure}(c), bonding distances in Table \ref{table:BondingDistance}. Although the van der Waals radii of the species in the ZL are smaller for the EMLG:N system ($1.70$~\AA\ for C vs.\ $1.92$~\AA\ for B), the distance of the EMLG:N layer itself to the ZL is clearly larger, even in absolute units ($3.28$~\AA\, for the \lqGnul layer it is  vs.\ $3.02$~\AA). This difference, $90.6$\% vs.\ $83.4$\% of the vdW distances, indicates a strong interaction of the \lqGnul layer with its substrate, in agreement with weak and broad Dirac bands seen in ARPES and faint reflection spots in LEED, while the decoupling of the \blue{EMLG:N} layer from the substrate is significantly better. 

Another obvious conclusion to be drawn from the analysis of bonding distances is that the \BNnul layer is located closer to the ZL than a  mere van der Waals interaction would suggest. \BBN and \NBN are located at $77$\% and $84$\% of the B-B and N-B van der Waals bonding distances, respectively, see Table \ref{table:BondingDistance}. This can be understood as a first indication for an (at least partly) chemisorptive (covalent) interlayer interaction.

\subsection{Calculation of the vertical structure by density functional theory}\label{sec:DFT}

\begin{figure}[b]
	\includegraphics[width=0.7\linewidth]{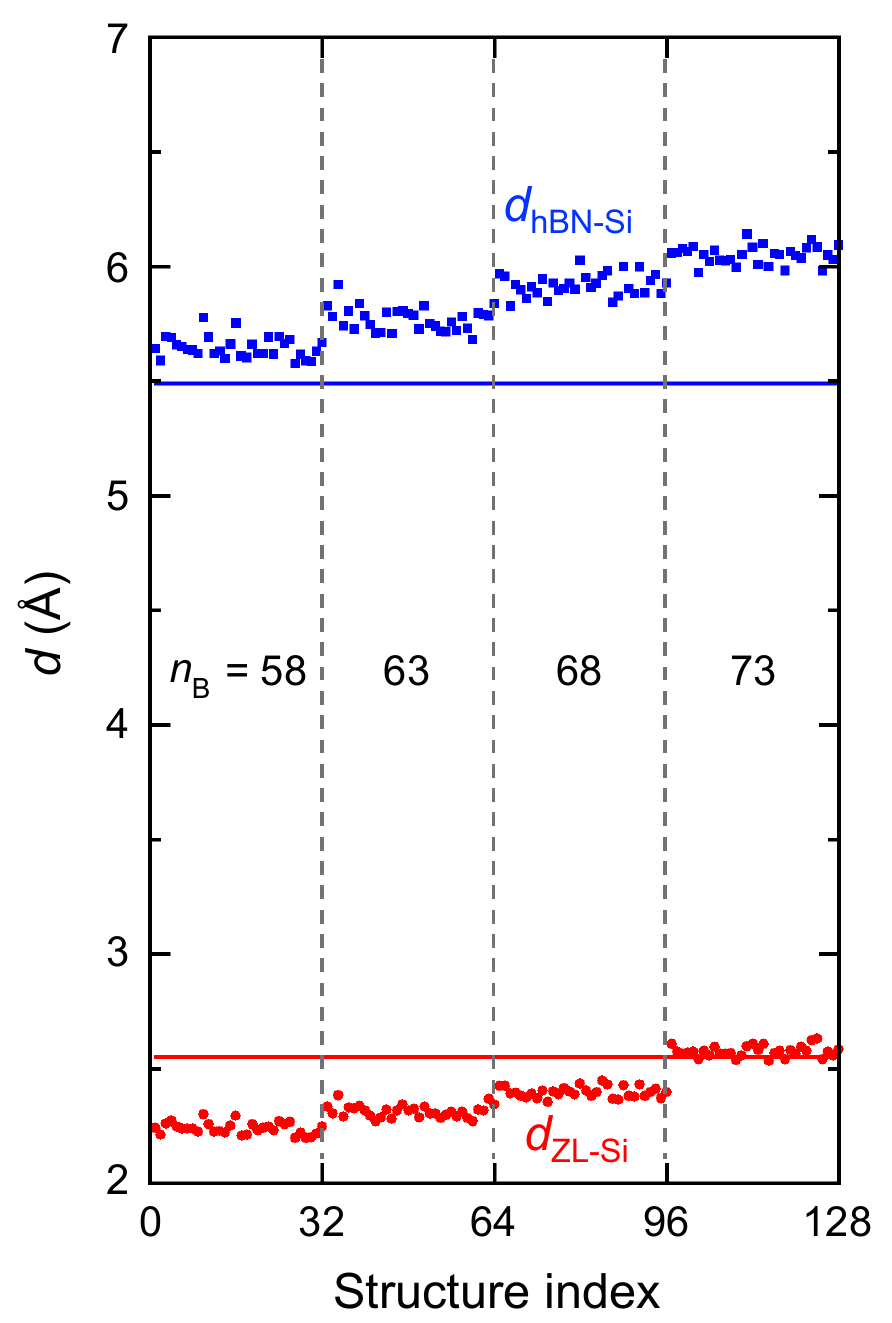}
	\caption{Interlayer distances $d_\text{hBN-Si}$ and $d_\text{ZL-Si}$ for 6H-SiC substrate + boron ZL + $(6\times 6)$-$R0^{\circ}$ hBN, predicted by DFT. As the structure of the ZL is unknown, it is modeled by randomly placing 58, 63, 68, or 73 boron atoms per supercell in a plane. 32 structures are studied for each number of boron atoms. Horizontal lines indicate the heights measured experimentally for the \BNnul sample prepared at 1150°C.}
	\label{fig:dft1}
\end{figure}

\begin{figure*}[t]
	\includegraphics[width=0.8\linewidth]{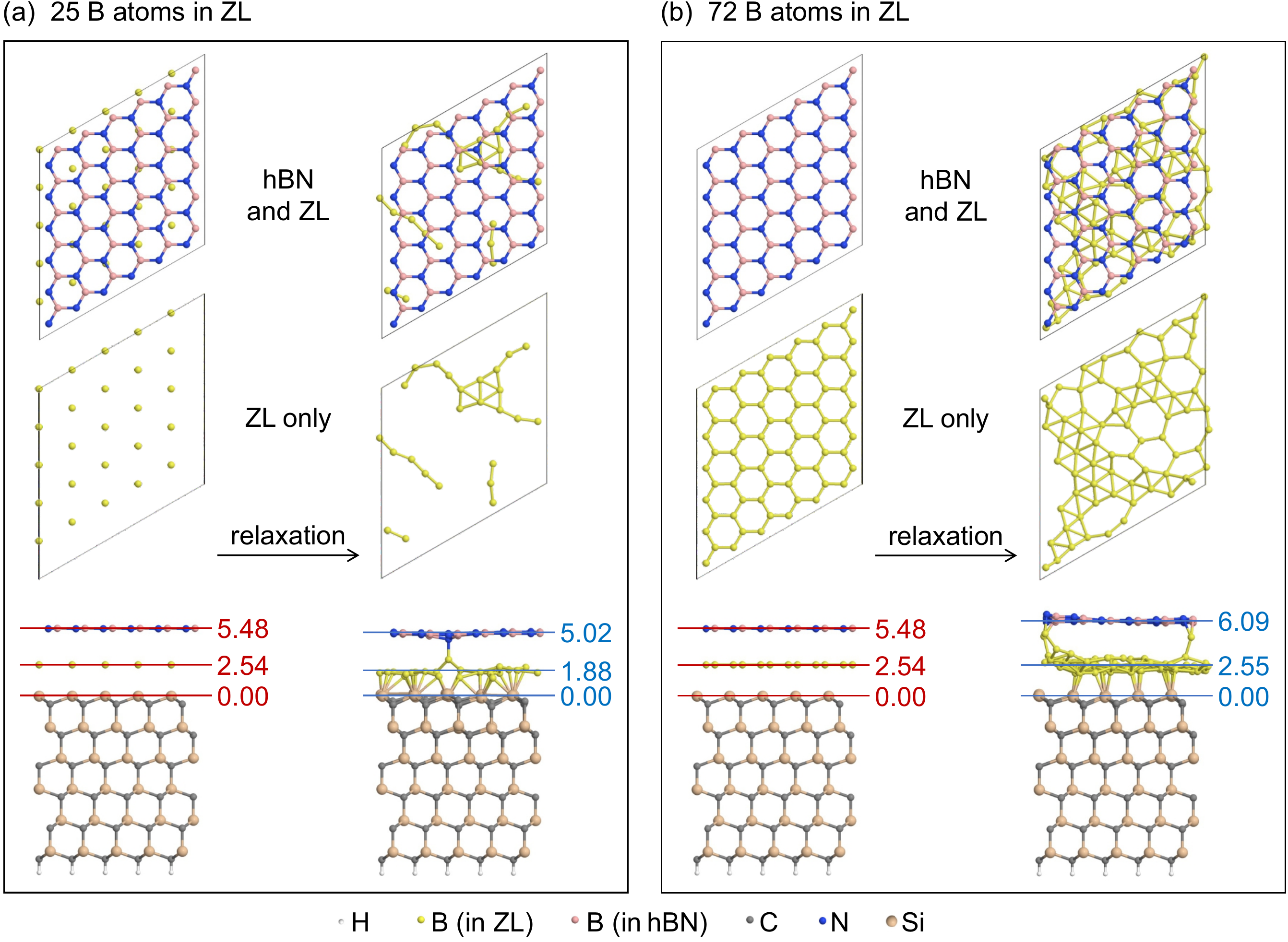}
	\caption{Top and side views of two model structures (6H-SiC substrate + boron ZL + $(6\times 6)$-$R0^{\circ}$ hBN), for which additional DFT calculations were caried out, before (left) and after (right) the geometry relaxation. The top view of the boron ZL alone is also shown. In (a), the ZL contains 25 B atoms, initially attached to Si atoms of the topmost SiC layer. In (b), the ZL contains 72 atoms, arranged in a honeycomb fashion. The vertical interlayer distances are shown in \AA ngstrom next to the layers.}
	\label{fig:dft2}
\end{figure*}

We have carried out DFT calculations with the aim to better understand the layer distances obtained by NIXSW.
However, the lateral atomic structures of the \BNnul layer as well as the ZL remain unclear, which makes it difficult to select the correct starting model for the DFT calculations. We therefore modeled a perfect 2D hBN layer ($x=y=1$), located above a boron ZL with random structure on the SiC(0001) substrate. Specifically, on a $5 \times 5 \times 1$ slab model of 6H-SiC we positioned the 2D hBN layer in two different supercells, namely $(6\times 6)$-$R0^{\circ}$ and $(\sqrt{39}\times \sqrt{39})$-$R16.1^{\circ}$. These supercells yield a biaxial strain in the hBN of $+2.56$\% and $-1.46$\%, respectively. Between the hBN layer and the substrate, a boron ZL is placed. Since we cannot be sure of the density $n_\mathrm{B}$ and the actual arrangements of the boron atoms in the ZL, we sample a group of random initial structures consistent with the overall features of the experiment: The simulated ZLs have densities $n_\mathrm{B}$ of  58, 63, 68, and 73 atoms per supercell, in altogether 128 randomly-generated structures. For the starting model, we took the vertical layer distances obtained by NIXSW (see Sec.~\ref{sec:NIXSW}). In the following, we discuss the $(6\times 6)$-$R0^{\circ}$-based structure models only, since they show the experimentally observed rotational alignment between substrate and overlayer. The results obtained with the $(\sqrt{39}\times \sqrt{39})$-$R16.1^{\circ}$ models lead to the same conclusion regarding interlayer distances.    

During the relaxation, only the bottom-most silicon, carbon and hydrogen atoms were kept fixed. The computed interlayer distances from the hBN layer and the boron ZL to the topmost plane of silicon atoms, denoted $d_\text{hBN-Si}$ and $d_\text{ZL-Si}$, respectively, are reported in Fig.\ \ref{fig:dft1}. Depending on the number of boron atoms in the ZL, $d_\text{ZL-Si}$ varies from $2.2$ to $2.65$~\AA.  We find that $d_\text{ZL-Si}$ increases as more boron atoms are added to the ZL. 
For the highest value of $n_\mathrm{B}$, 73 atoms per supercell, the best agreement of simulated and measured values is obtained, $d_\text{ZL-Si} = 2.65$~\AA\ in the simulation,  $2.54$~\AA\ in the experiment. Hence, based on our simulations and independent of the actual local ZL structure, we propose a boron density of approx.\ 73 atoms per supercell for the ZL. However, the calculations for this high boron density also show that the calculated height of the hBN layer ($d_\text{hBN-Si} = 6.15$~\AA) does not agree with the correponding experimental value of $5.50$~\AA. This is because the calculated hBN-ZL distance ($3.4$ to $3.6$~\AA) never comes close to the measured value for the \BN-ZL distance of $2.96$~\AA\ (see Fig.\ \ref{fig:dft1}), and clearly indicates that the perfect-hBN model does not provide a correct description of the experimental \BNnul layer.

In addition to the randomly-generated structures, we have tested two boron ZL models as shown in Fig.\ \ref{fig:dft2}. In the first model, the ZL has as many boron atoms as silicon atoms in the topmost SiC layer, that is a density of $n_\mathrm{B}=25$ atoms per supercell. Each boron atom is attached to a silicon atom. The computed $d_\text{ZL-Si}$ is 1.88 \AA, significantly lower than the experimental value. The second model is a uniform hexagonal sheet of boron atoms, resulting in $n_\mathrm{B}=72$ atoms per supercell. The computed $d_\text{ZL-Si}$ is 2.55~\AA, in excellent agreement with the experiment (2.54~\AA). However, as in the simulation of randomly generated structures for the ZL, with $6$~\AA, $d_\text{hBN-Si}$ is much too high here as well.

We conclude that the DFT calculations can reliably provide the density of boron in the ZL by simulating many randomly-generated structures and comparing simulated and experimental ZL-Si distances. Furthermore, the simple model of perfect hBN positioned above the ZL (with correct density) cannot reproduce the experimental vertical structure in terms of \BN-ZL and \BN-Si distances. Hence, the calculations indicate that the layer structure found experimentally for the \BNnul sample is not compatible with 2D hBN on 6H-SiC(0001).

\begin{figure}[b]
	\includegraphics[width=0.82\linewidth]{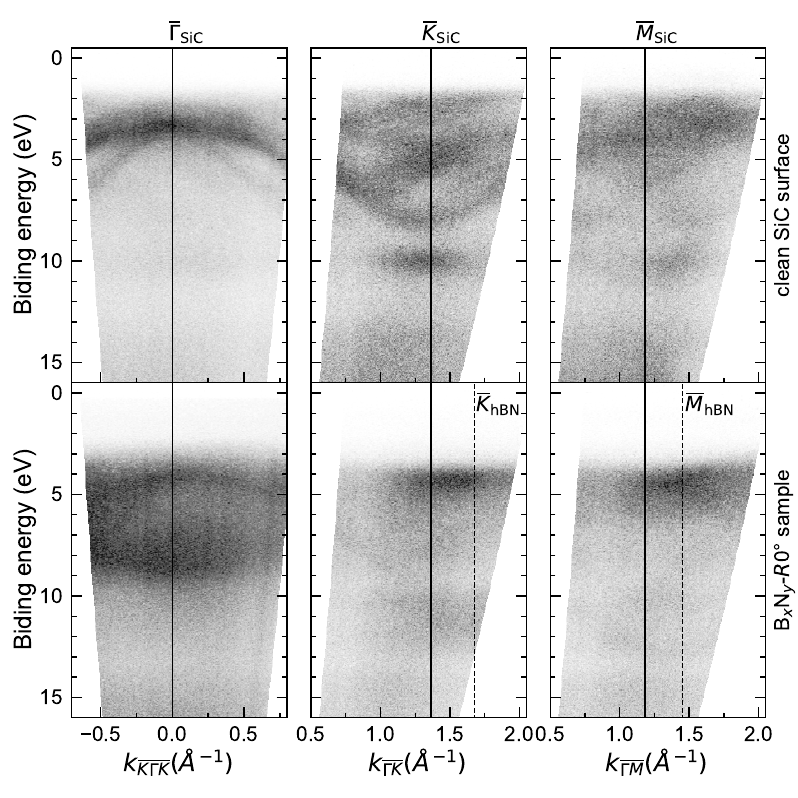}
	\caption{\label{fig:ARPES_BN} Band maps of the clean \three reconstructed SiC surface (upper) and of the \BNnul sample prepared at 1150\C (lower). $h\nu=40.8$~eV (He II).}
\end{figure}

\subsection{Electronic structure and air stability}\label{sec:electronic+air}

Finally, we have performed ARPES measurements in order to help identify the band structure of the layers formed on the SiC surface at different preparation temperatures. Fig.\ \ref{fig:ARPES_BN} shows band maps around the $\overline{\Gamma}$, $\overline{K}$ and $\overline{M}$ points of the clean SiC(0001) surface (upper part) and of \BNnul sample, prepared at $1150$\C in borazine atmosphere (lower part). The data were recorded using He II radiation ($h\nu = 40.8$~eV).

The maps taken on the \BNnul sample exhibit some broad and faint bands, but no indications of the typical band structure of decoupled 2D hBN \cite{Catellani1987, Auwaerter2019}. This finding is in contrast to Ref.\ \cite{Shin2015}, but was confirmed by repeating the experiment using He I ultraviolet light and soft x-ray synchrotron radiation ($h\nu = 21$~eV and $h\nu = 110$~eV, respectively, data not shown), with the same result. Note that in the SPA-LEED measurements we found an average domain size of 30~nm for the \BNnul layer (see Sec.~\ref{sec:SPA}), a size that would be sufficient to provide a clear band structure in ARPES, if the structure was hBN. 

\begin{figure}[t]
	\includegraphics[width=\linewidth]{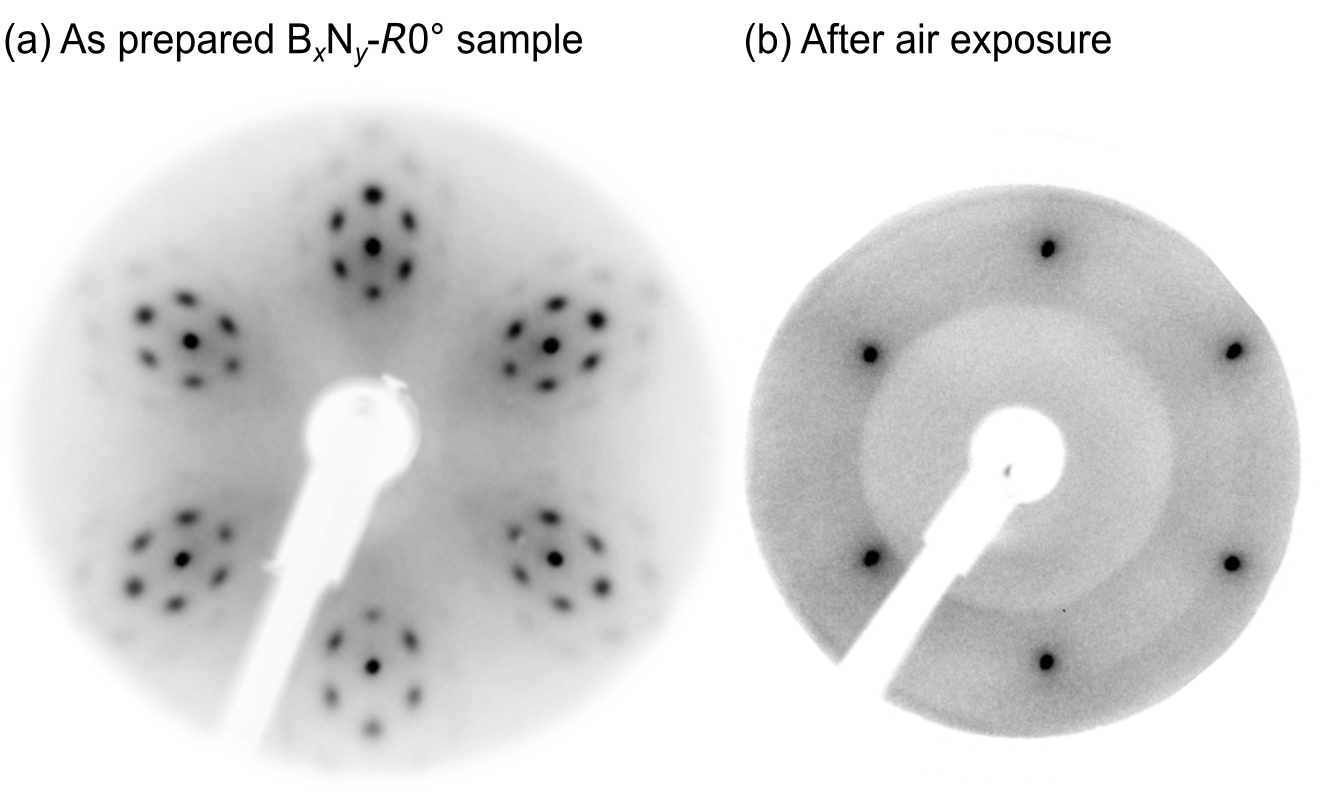}
	\caption{\label{fig:airStab} Diffraction patterns of a \BNnul sample (a) before ($E=100$~eV) and (b) after air exposure for 48 hours ($E=110$~eV). All LEED spots have vanished after air exposure, except those of the substrate.}
\end{figure}

\begin{figure}[t]
	\includegraphics[width=\linewidth]{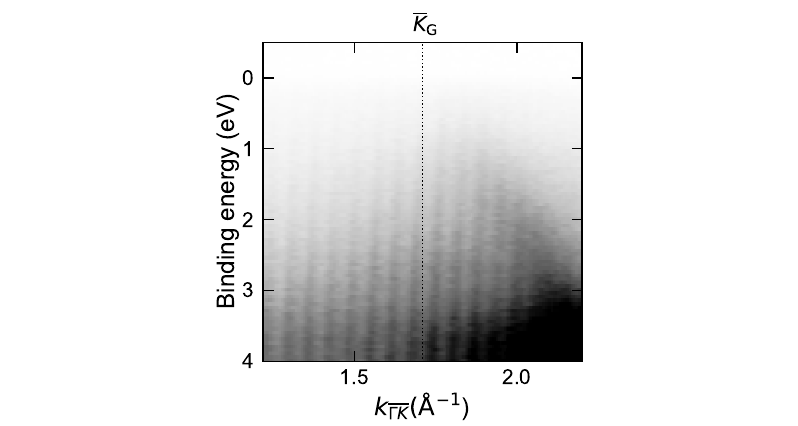}
	\caption{\label{fig:ARPES_Gr} Band map of the \lqGnul sample prepared at 1225\C. $h\nu=110$~eV.}
\end{figure}

The same conclusion -- the \BNnul layer is not a decoupled 2D hBN layer -- can be drawn from diffraction experiments performed on samples which were transferred in air: Figure \ref{fig:airStab} shows a comparison of LEED patterns recorded from a \BNnul sample directly after preparation, and after 48 hours exposure to air. Even with mild annealing in UHV, the original pattern cannot be recovered. hBN, however, is known to be stable in air \cite{Auwaerter2019, Liu2013, Rigosi2021}. 

In contrast, for samples prepared using the same procedure but higher temperatures ($1225$\C), a faint and broad Dirac cone was found in the energy dispersion map, as displayed in Fig.\ \ref{fig:ARPES_Gr}. This confirms that the \lqGnul layer starts to emerge from the \BNnul layer. 

\section{Conclusion}

We investigated the formation of boron nitride (\BNnul) and graphene (\lqGnul) layers on 6H-SiC(0001). The layers are formed by annealing the SiC bulk crystal in borazine atmosphere at temperatures between $1100$\C and $1250$\C atop a boron ZL. The geometric and electronic structure of both layers, as well as the transformation of \BNnul to \lqGnul were investigated.

Our main conclusion is that -- in contrast to the existing literature \cite{Shin2015} -- hBN does not stabilize on a SiC(0001) surface, at least not when using the here employed preparation protocol, which does not appear to differ significantly from the one reported in Ref.\ \cite{Shin2015}. Results from several complementary methods unambiguously indicate that the \BNnul layer forming at $1100$\C-$1150$\C is not a 2D hBN layer: Although it has a hexagonal structure and is aligned with the substrate, it is not decoupled from the surface. NIXSW revealed bonding distances that are $\approx 20\%$ smaller than van der Waals distances and not compatible with DFT calculations for a perfect 2D hBN layer. In ARPES, the \BNnul layer does not show the typical band structure of hBN, and it is not stable in air.

For the boron ZL at the interface between \BNnul and SiC we were able to determine the density. It is close to a hypothetical uniform hexagonal layer of boron atoms, and hence likely lower than that of the borophene structures proposed in Cuxart \etal and Hou \etal \cite{Cuxart2021, Hou2020}. Furthermore, in electron diffraction and by DFT-based structure simulation on a large number of initial guesses, the ZL layer does not exhibit any long-range order.

At preparation temperatures higher than 1225\C, the \BNnul layer transforms gradually to the \lqGnul layer conserving its orientation. A certain amount of nitrogen remains in the layer (as indicated by the almost identical adsorption heights for C and N). This high doping level might contribute to the poor quality of the unconventionally oriented \lqGnul layer. 
The interaction of the \lqGnul layer with the underlying substrate is also relatively strong, much stronger than for the case of an \blue{EMLG:N} layer (bonding distances of $83\%$ and $90$\% of the expected vdW distances, respectively).

Note that the quality of the \lqGnul layer can be significantly improved by using the so-called ``surfactant method'' for preparation, as proposed in our previous work \cite{Bocquet2020}. Following this preparation route, which basically applies an even higher preparation temperature, no \BNnul structure is formed prior to the formation of a high quality $R0^\circ$ graphene layer that is decoupled from the SiC substrate by a graphene ZL. This system can serve as a perfect starting point to produce a $30^\circ$ twisted bilayer graphene by transforming the carbon ZL into a (conventionally oriented) graphene layer, e.g., by hydrogen intercalation.

All experimental data shown in the main text are available at the Jülich DATA public repository \cite{DATA-repository}. All simulation data shown in the main text are available at the Nomad public repository \cite{Sim-DATA-repository}.

\section*{acknowledgments}
We thank Diamond Light Source for access to beamline I09 (via proposal SI-17737), and the I09 beam-line staff (P.~K.\ Thakur, D.\ Duncan, and D.\ McCue) for their support during the synchrotron experiment. We are also grateful for the support by Nafiseh~Samiseresht during the experiments. {Y.-R.~L.}, {F.~C.~B.}, {C.~K.} and {F.~S.~T.} acknowledge funding by the Deutsche Forschungsgemeinschaft (DFG, German Research Foundation) through SFB 1083 ``Structure and Dynamics of Internal Interfaces'', sub-project A12. 
V. W.-z. Y. and V.~B. were supported by the National Science Foundation under Award No. ACI-1450280. An award of computer time was provided by the INCITE program. This research used resources of the Argonne Leadership Computing Facility, which is a DOE Office of Science User Facility supported under Contract DE-AC02-06CH11357.

\section*{Author contributions}
F.~S.~T., C.~K. and F.~C.~B. conceived the research.
Y.-R.~L., M.~F. and F.~C.~B. prepared the samples. 
Y.-R.~L., M.~F., S.~P., M.~R., T.-L.~L., S.~S., C.~K. and F.~C.~B. performed the XPS and NIXSW experiments at Diamond Light Source, 
Y.-R.~L. and S.~P. performed the SPA-LEED experiments, 
and Y.-R.~L. performed the ARPES experiments.
Y.-R.~L. analyzed all experimental data and made the corresponding figures. 
V. W.-z. Y. and V. B. planned the DFT calculations. V. W.-z. Y. carried out all DFT calculations and made the figures showing the DFT results.
Y.-R.~L., C.~K. and F.~C.~B. wrote the paper, with significant contributions from V. W.-z. Y., V.~B., and F.~S.~T.

\end{document}